# Towards Random Metasurface based Devices

**MATTHIEU DUPRÉ, LIYI HSU, AND BOUBACAR KANTÉ**[*]

*University of California San Diego, Department of Electrical and Computer Engineering, 9500 Gilman Drive, La Jolla, CA 92093, USA*
*\*bkante@ucsd.edu*

**Abstract:** Metasurfaces are generally designed by placing scatterers in a periodic grid. We propose and discuss design rules for efficient random and functional metasurfaces with anisotropic elements. We investigate the impact of the elements density on the performance of metasurfaces in both periodic and random cases. Using numerical simulations, we apply our approach to the design of all-polarization random metalenses at near infrared frequencies. The results will open a new paradigm in the design of metamaterial and metasurface devices from lenses to solar energy concentrators.

Metamaterials, on one hand, are artificial materials designed at the subwavelength scale, with exotic behavior, from negative refraction [1,2] to cloaking [3] through subwavelength focusing [1]. However, their intrinsic absorption has limited their practical implementation for real life applications; even if such dissipation can be mitigated [4]. On the other hand, metasurfaces, the two-dimensional version of metamaterials, do not suffer such propagation losses. Originally designed for radio waves for radar and space communications [5,6], metasurfaces have been used to design devices at visible and infrared wavelengths such as carpet cloaking [7,8], holograms [9], optical lenses [10] and solar concentrators [11] to name a few.

Metasurfaces control the waves reflection and refraction at interfaces using phase-shifting elements [12]. In optics, whether they are designed from metallic materials using plasmonics phenomena or dielectrics to obtain higher efficiencies at the cost of larger elements, whether they are relying on resonators, waveguides or geometric phase to tune the phase of the wave, metasurfaces are generally designed in a periodic framework where their constituting elements are placed on a periodic grid [13]. The relation providing the phase-shift of the elements as a function of their dimension is calculated either analytically, when possible, or with numerical simulations for a single element or for periodic arrays of identical elements. However, metasurfaces are generally made of elements of different sizes to provide a phase-shift that varies spatially. Hence, the previous approaches may fail as near-field coupling introduces errors in the phase-shifts provided by the elements. We can therefore wonder whether this periodic arrangement is the best solution and whether we can design metasurfaces within a random framework. Another major advantage of random or disordered materials is their statistical isotropy [14]. This translates to the metasurface world as polarization independent metasurfaces, even for very anisotropic particles that provide polarization dependent phase-shifts. A few studies took advantage of this property and used random metasurfaces for reducing the radar cross-section [15–18]. However, more diverse devices still need to be realized as their designs remain elusive due to the disordered distances between neighboring elements, the near-field coupling, and variations of the local density of elements. Some theoretical approaches can address the homogenization problem of homogeneous random polarizability materials in periodic arrays of resonators [19], or for identical polarizabilities in disordered arrays of scatterers [20,21]. However, more work is required to design advanced metasurfaces with phase gradient.

Here, we adopt a more practical approach and propose to use anisotropic gold nanoparticles resonators to design random metasurface lenses at a wavelength $\lambda_0$=1.5 μm. We start by considering 1D periodic metasurfaces. Using numerical simulations, we investigate the effect of the density of elements on the metasurfaces. This allows us to identify optimal densities for

1D random metasurfaces. Finally, we extrapolate our approach to the two-dimensional case to design 2D polarization independent random metalenses.

Fig. 1(a) shows the unit cell of the metasurfaces that is considered in this paper. It is made of a gold nanoparticle of width 50 nm, height 40 nm, and the length is tuned between 150 and 500 nm. Gold is modeled using a Drude model with a plasma frequency $\omega_p=1.367\times10^{16}$ rad/s and a collision frequency $\omega_c=6.478\times10^{13}$ rad/s [22]. This plasmonic particle is supported by a dielectric SU8 spacer with a refractive index $n_{SU8}=1.5875$, of thickness 100 nm, on top of a metallic ground plane, that is modeled as a perfect electric conductor (PEC) in our simulations, and which can be easily made of gold or silver for practical implementations.

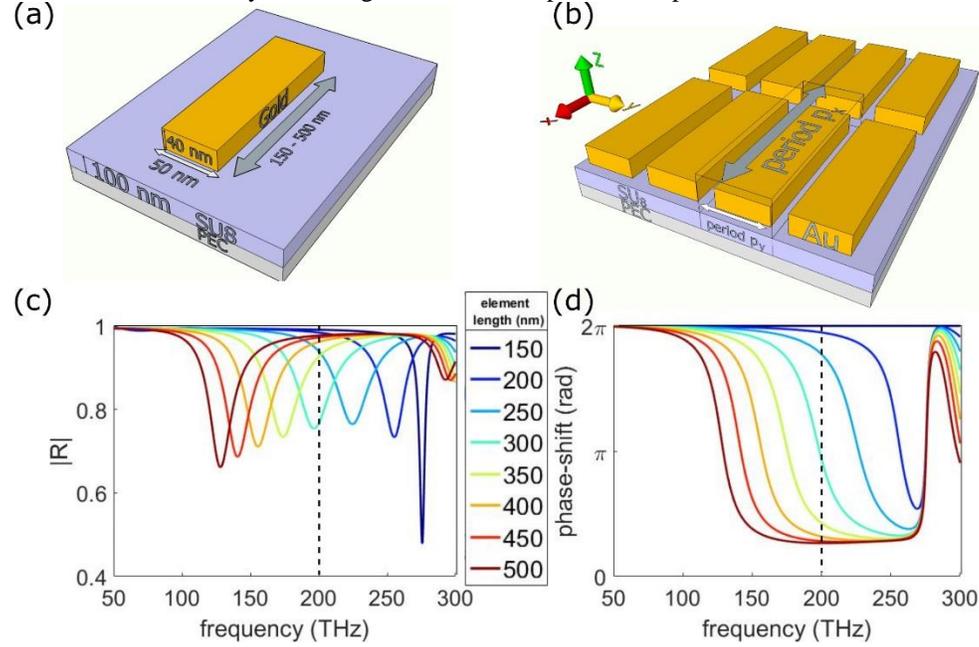

Fig. 1(a) Unit cell schematic of the metasurface. The gold nano-elements dimensions are: 50 nm width, 40 nm height, and a length varying from 150 to 500 nm. The dielectric spacer, SU8, has a thickness of 100 nm and is on a metallic (PEC) substrate. (b) An array of particles with periodic boundary conditions. (c) Amplitude of the reflection coefficient as a function of the frequency around 200 THz and for element lengths from 150 nm to 500 nm. Vertical dashes mark the 200 THz frequency. $p_x=900$ nm and $p_y=150$ nm. (d) Same as (c) for the phase-shift of the reflection coefficient.

We first investigate such elements using in-plane periodic boundary conditions, as shown on Fig. 1(b). The period along the width of the element (period $p_y$ in the $y$ direction) will be optimized but is kept to 150 nm in Fig. 1. The period in the longer dimension of the element ($p_x$ in the $x$ direction) is set to 900 nm and is kept constant all along the paper. Using the frequency domain solver of the commercial software CST, the reflection coefficient of our structure is computed. The illumination is a plane wave with a frequency varying from 50 THz to 350 THz, polarized along $x$, (i.e. the long axis of the particle). The particle is transparent to the orthogonal polarization. Varying the length of the particle from 150 to 500 nm shifts its fundamental frequency as shown on Fig. 1(c). The phase of the wave reflected by the particle and the metallic plane can thus be controlled. Fig. 1(d) shows the phase shift of an element around 200 THz—which corresponds to a wavelength of $\lambda_0=1500$ nm—for different particle lengths. The shortest element is taken as phase reference which explains the apparent increase of the phase of any other element at frequencies higher than 275 THz. Fig. 2(a) shows the phase shift as a function of the length of the nano-bars for different spacer thicknesses at 200 THz. For a single resonance, the complete $2\pi$ phase shift is only obtainable asymptotically far away from the resonance. The SU8 spacer thickness allows us to optimize the quality factor Q of the resonances which in turn, controls the maximum value of the phase-shift —the thinner the SU8 layer, the lower the Q factor and the smaller the maximum phase-shift. However, the higher the Q factor, the sharper the slope of the reflected phase. Hence, a compromise has to be made between the maximum value of the phase shift and the slope of the reflected phase around 200 THz (Fig. 1(d)). Indeed, a very steep change of phase introduces discretization errors. A thickness of 100 nm (the green curve on Fig. 2(a)) seems to be a good compromise as the difference between $2\pi$ and the maximum phase-shift is small and inferior to 35°. This is a tolerable error for a metasurface, while the slope of the curve is small enough to enable discretization every 10 nm of length.

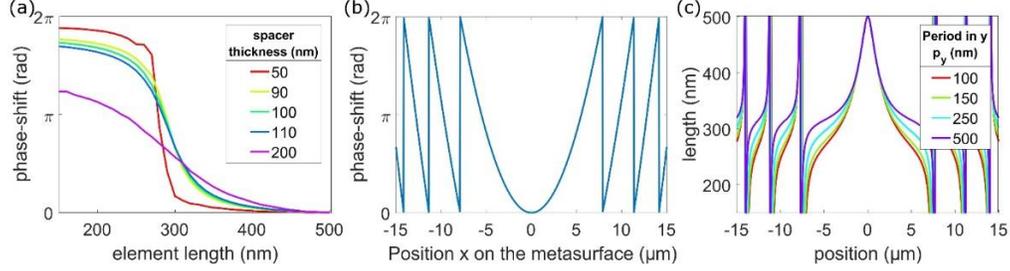

Fig. 2. (a) Phase shift provided by an element at 200 THz as a function of the length of the nano-bar, for different SU8 layer thickness. (b) Phase shift required as a function of the position, to design a lens of 30 μm width with a focal length of 20 μm at 200THz. (c) Length of the elements as a function of the position to design the lens. Curves are obtained from (a) and (b) for different periods in the y (short dimension of the elements) direction.

The phase-shift required to design a lens or concentrator with a focal length $f$ is given by the parabolic law:

$$\varphi(x) = k_0\left(\sqrt{x^2 + f^2} - C\right) \qquad (1)$$

where $C$ is an arbitrary constant generally chosen to be equal to $f$ to have a null phase shift at the center of the lens. The phase-shift required to design a metalens of 30 μm width with a focal spot of 20 μm is shown on Fig. 2(b). Knowing the phase-shift required at any position of the metasurface (Fig. 2(b)) and the phase shift induced at reflection as a function of the nano-bar length (Fig. 2(a)), we can directly obtain the length of the nano-bars as a function of their position on the metasurface. Fig. 2(c) presents the length required at a given position to realize the phase shift plotted in Fig. 2(b), for different periods from 150 nm to 500 nm. Changing the period shifts the resonant frequency of an array of identical elements. This originates from two reasons: near-field coupling that becomes stronger as the distance between the particles is decreased, or the density of particle itself. Indeed, the denser the particles, the more field will be phase-shifted by the particles compared to the field which is only reflected by the ground plane. The total reflected field which is the sum of the field reflected by the mirror and the field scattered by the elements has therefore different phases for different densities.

Knowing the length of the elements at any position, we design periodic one-dimensional metasurfaces. We have to distinguish two types of period. The first one is the period that we choose to compute the reference providing the phase as a function of the length of the elements, in a periodic environment as in Fig. 2(a). We design several metasurfaces for different periods with references of 100, 150, 250 and 500 nm and that correspond respectively to densities of 11.1, 7.5, 4.4 and 2.2 elements per μm², or to 4.9 $\lambda_0^{-2}$, 3.3 $\lambda_0^{-2}$, 2 $\lambda_0^{-2}$ and 1 $\lambda_0^{-2}$. We define the density here as $\rho = 1/(p_x p_y) = N/(p_x L_y)$, where N is the number of elements, $L_y$ the length of the metasurface and $p_x$ and $p_y$ the periods in the two dimensions. The second type of period is nothing else but the real period of the metasurface. Indeed, when we design a metasurface, we can choose independently the real period from the period that we used to plot the curve of Fig. 2(a). This comes to change the density of elements on the metasurface. In other words, the reference period (or the density of reference) serves as a reference to compute the length of the element at a given position, while the real period (or density) is used to set the density of the metasurface. Using CST simulations in time domain for better computational efficiency, we compute the density of energy of the reflected field normalized by the density of energy of the incident plane wave for 1D periodic metasurfaces designed for the four period of the reference times the four real periods and at the nominal design frequency of 200 THz, λ=1.5 um. Results are presented in Fig. 3.

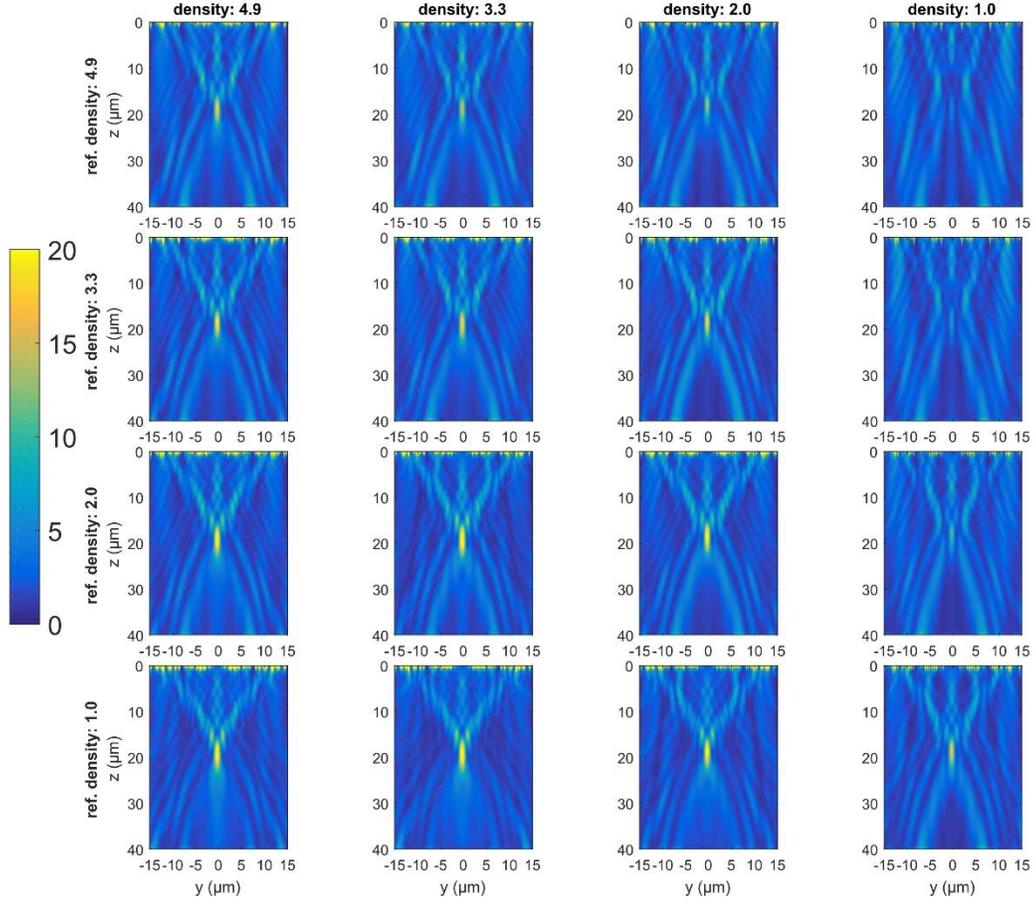

Figure 3: 1D periodic metasurfaces. Density of energy of the reflected field normalized by the density of energy of the incident fields as a function of the y and z (out of plane) positions, for 16 one-dimensional periodic metalenses: for four different periods of the reference and four different periods of the metasurface. The different considered periods $p_y$ are 100, 150, 250 and 500 nm and correspond respectively to 4.9, 3.3, 2 and 1 elements per wavelength squared.

All the field plots show focal spots of various intensities. The focal spots dimensions (full width at half maximum) are about 1.3 μm wide and 6 μm long. For comparison, the Abbe limit for a lens with a numerical aperture of 0.6 is 1.25 μm by 4.2 μm. The results on the diagonal provide the best performances, as these are the metasurfaces designed with the same densities as the periodic references. We note that the metasurfaces situated in the upper part of the figure provide poorer performances. On the contrary, the metasurfaces situated below the diagonal provide performances almost as high as the diagonal elements. This may seem surprising but the error of the phase-shift caused by the period mismatch between the metasurfaces and their reference is compensated by the higher number of elements which provide a higher amplitude to the reflected field.

We then conduct similar simulations for 1D random metasurfaces. The approach to design a random metasurface is the following. First, we randomly select a position (for the 2D cases, we also choose a random orientation for the element). We compute, using the reference curves (Fig. 2(a-c)), the length that this particle should have. Then, we check if it is overlapping or is too close to previously placed particles. We set this minimum distance to be 10 nm in order to put the particles as close as possible to get the maximum density. If the particles are too close or overlap, we remove it and select another random position. If they do not overlap, we approve the change and move to the next particle. We repeat the process until we manage to place a defined number of elements (from 300 for a density of 4.9 $\lambda_0^{-2}$ to 60 for a density of 1 $\lambda_0^{-2}$) or until we have failed a certain number of time to place a given element. In the last case, the maximal density of the random metasurface has been reached. The maximum density of 4.9 $\lambda_0^{-2}$ is low enough to ensure we can place all the elements on the metasurface. As we are working with random metasurface, we repeat the process and simulate each design ten times to ensure that the results are statistically significant. The average results for the 16 (4 by 4) metasurfaces are displayed on Fig. 4. Unaveraged results look very similar as the maximum value of the standard deviation is about 10% of the average value. The densities and densities of reference correspond to the cases of periodic metasurfaces. Hence, a density of reference of 4.9 elements per wavelength squared corresponds to a period of 100 nm and to the curve plotted in red in Fig. 2(c). The frequency is still 200 THz which corresponds to a wavelength of 1.5 µm. All elements are kept parallel to the polarization direction of the incident field.

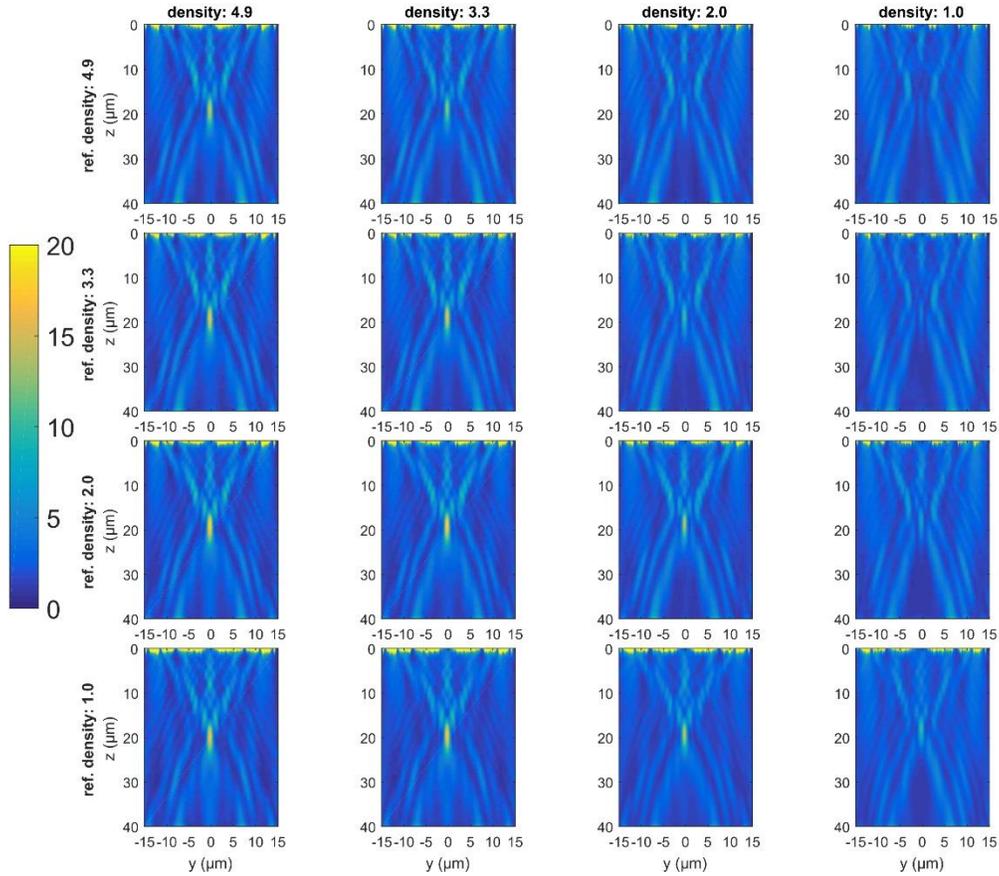

Figure 4: 10 times average of 1D random metasurfaces. Density of energy of the reflected field normalized by the density of energy of the incident fields as a function of the y and z (out of plane) axis, for 16 one-dimensional random metalenses: for four different densities of the reference times four different densities of the metasurface. The different considered densities are 4.9, 3.3, 2 and 1 $\lambda_0^{-2}$, and correspond to periods of 100, 150, 250 and 500 nm in the periodic case.

In Fig. 4, we can observe similar characteristics as in Fig. 3 between low performance metasurfaces in the upper right and better performance for the bottom left elements. Here, we also observe better results for the highest densities, in the first two columns. The performance of such metasurface is close to periodic metasurfaces. The period of the reference does not play such a major role as in the periodic case because the distances between elements is not constant due to the inherent disorder of random metasurfaces.

Hence, having densities as high as possible is of prime importance to design random metamaterials in order to compensate for the phase shift mismatch brought by the intrinsic disorder compared to the references.

In what follows, we now consider 2D metalenses. Metasurface are based on the reference curves corresponding to a period of 100 nm. This is quite close to the maximum density achievable for 2D random metasurfaces for our rectangular elements. To limit the required computational power, we decrease the size of the metasurfaces to 10 by 10 μm, with a focal length of 10 μm. The nominal frequency is still 200 THz for a wavelength of $\lambda_0=1.5$ μm. The first metasurface is a triangular lattice metasurface—i.e. periodic—made of 1121 vertically aligned elements (Fig. 5(a)), which corresponds to a density of 11.2 μm$^{-2}$ or 5 $\lambda_0^{-2}$ (5 elements per wavelength squared). The triangular lattice allows higher densities of elements compared to a rectangular one. The second metasurface is made of the same number of elements but with randomly placed and oriented elements (Fig. 5(b)). We also impose a minimum distance of 10 nm between adjacent elements in the random metasurfaces. An interesting feature of random metalenses is visible in Fig. 5(b): the density of element is not homogeneous. At locations where the elements are the larger, the density is smaller, while positions where elements are shorter, the density is higher. We can expect such distribution to compensate the fact that smaller elements generally have a smaller scattering cross-section. The density of energy of the reflected waves is plotted in Fig. 5(c,d). We can see that the two metalenses focus light at a distance slightly inferior to the focal length, 8 μm instead of 10 μm, due to the short focal length and small size of the metasurface compared to the wavelength. However, the random metalens (Fig. 5(d)) has a much lower efficiency (about 15% at the maximum of the focal spot) compared to the periodic metasurface which is around 75%. Intuitively, one would expect the efficiency of the random metasurface to be half the value of the periodic metasurface due to its polarization properties. Indeed, the random metalens can focus both polarization due to its randomly orientated elements, while the resonators of the periodic metasurface are all aligned. Hence, the latter can only focus one polarization. We attribute this lower efficiency to the near-field coupling that is different for the random metasurface compared to the 1D case where all the elements were aligned. If we increase the frequency of the incident plane wave, the efficiency of the periodic metasurface drops while the efficiency of the random one increases. Fig. 5(e) presents the density of energy for the random metasurface at 240 THz. We can see that the efficiency has almost doubled (and reaches a value of 27.8%) compared to 200 THz (Fig. 5(d)) and has a much lower relative background. The maximum of the focal spot also moved closer to the nominal position of the focal length (10 μm). Hence the disorder of orientation plays a crucial role in the near-field coupling that, in turn, controls the efficiency of the metasurface. Studying and optimizing 2D random metasurfaces is not as straightforward as 1D metasurfaces and greater details, beyond the scope of this work, will be presented in a forthcoming paper.

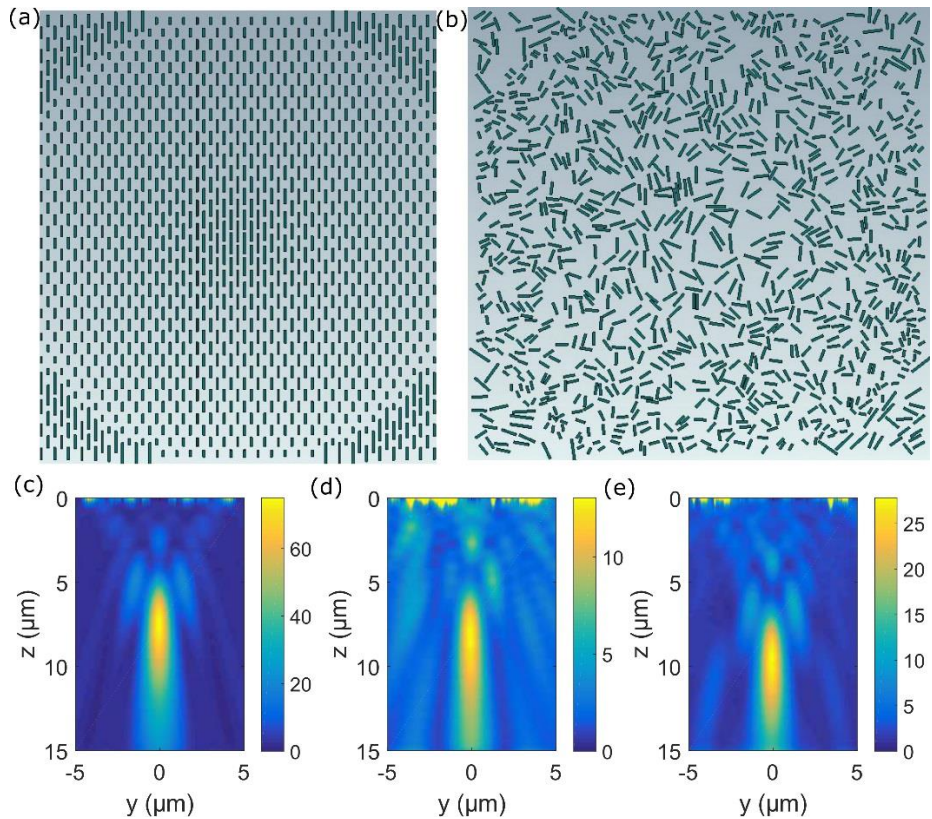

Figure 5: 2D metalens. (a) periodic metalens made of 1121 elements placed on a 10x10 um$^2$ triangular lattice. Focal length f=10 μm. Period of the reference = 100 nm. (b) 2D random metalens similar to (a) and with the same number of elements (1121) but with randomly placed and oriented elements. (c) Density of energy of the reflected field of the periodic metalens in (a) in the plane *x=0* at 200 THz. (d) Density of energy of the reflected field of the random metalens in (b) in the plane *x=0* at 200 THz. (e) Density of energy of the reflected field of the random metalens in (b) in the plane *x=0* at 240 THz.

We have shown, using numerical simulations, that 1D and 2D functional random metasurfaces can be designed. Our results indicate that the efficiencies of 1D random metalens can be as high as the efficiencies of periodic ones if the density of elements is high enough. We have shown that 2D random metalenses that successfully focus light of any polarization. The efficiency of random metalens mainly depends on the density of elements. Further works need to be performed to understand the role of the orientation disorder and the strength of the near-field coupling in order to optimize 2D random metalenses. Our results pave the way to the design of random metasurfaces for devices as diverse as lenses and concentrators. We also believe that random metasurfaces may overcome limitations on the diffraction efficiency of periodic systems.

This material is based upon work supported by the U.S. Department of Energy under Award Number DE-EE0007341.